\documentclass[10pt,a4paper]{article}
\usepackage{graphicx}
\usepackage{mathrsfs}
\usepackage{fancyhdr}
\usepackage[top=50truemm,bottom=60truemm,left=30truemm,right=30truemm]{geometry}
\usepackage{indentfirst}
\usepackage{authblk}

\renewenvironment{abstract}{\begin{quotation}{\normalsize }}{\end{quotation}}
\makeatletter
\newcommand{\email}[1]{\newcommand{\@email}{E-mail: #1}}
\renewcommand{\maketitle}{
\newpage\null
    \vspace{2em}
\begin{center}
 {\Large\bfseries\noindent\ignorespaces\@title\par}
 \vspace{1em}%
 {\large\noindent\ignorespaces\@author\par}
 \vspace{2mm}
 {\normalsize\noindent\ignorespaces\@email\par}
\end{center}
\vspace{1em}
}

\renewcommand\@author{\ifx\AB@affillist\AB@empty\AB@author\else
      \ifnum\value{affil}>\value{Maxaffil}\def\rlap##1{##1}%
     \vspace{1mm} \AB@authlist\\\AB@affillist
    \else  \AB@authors\fi\fi}
\makeatother
\usepackage{amsmath,amssymb}
\title{Present Status of Fission Research Based on TDDFT}
\author{Yoritaka Iwata}
\affil{Center for Nuclear Study, The University of Tokyo, Hongo 7-3-1, 113-0033 Tokyo, Japan}
\email{iwata@cns.s.u-tokyo.ac.jp}
\begin{document}
\maketitle
\begin{abstract}
Fission resulting from collision of atomic nuclei is systematically
investigated based on time-dependent density functional calculations. 
Time-dependent density functional theory (TDDFT) is a framework, which enables us to treat quantum many-body dynamics with nucleon degrees of freedom.
In this article a theoretical framework called ``Composite-Nucleus Constrained
TDDFT'' is introduced, and charge equilibrium hypothesis for collision fission
dynamics is examined.
\end{abstract}


\section{Introduction}
Fission of the nucleus is important to many processes.
For instance fission should play an important role in superheavy
synthesis, as well as many astrophysical phenomena.
As for the researches based on the TDDFT, there are several developments in the fission research recently (for example, see \cite{14wakhle,14oberacker,14simenel,15guillaume}).

Let us imagine the collision between nuclei.
We use a terminology collision-fission to denote the fission resulting from collision. 
Fusion-fission and quasi-fission are the main components of the collision-fission.
There are several stages in nuclear collisions depending on the time-scales \cite{13iwata}.
At the early stage there is a contact between the two nuclei, and the quite rapid processes ($\sim$ 10$^{-22}$s) such as the fast charge equilibration follows \cite{10iwata,12iwata}.
At the intermediate stage ($\sim$10$^{-21}$s) composite nucleus is formed, and the collective oscillation such as giant resonance follows.
Sometimes after a sufficient time ($> 10^{-20}$s), fission appears.

\begin{figure} [t]
\begin{center} 
\includegraphics[width=12.0cm]{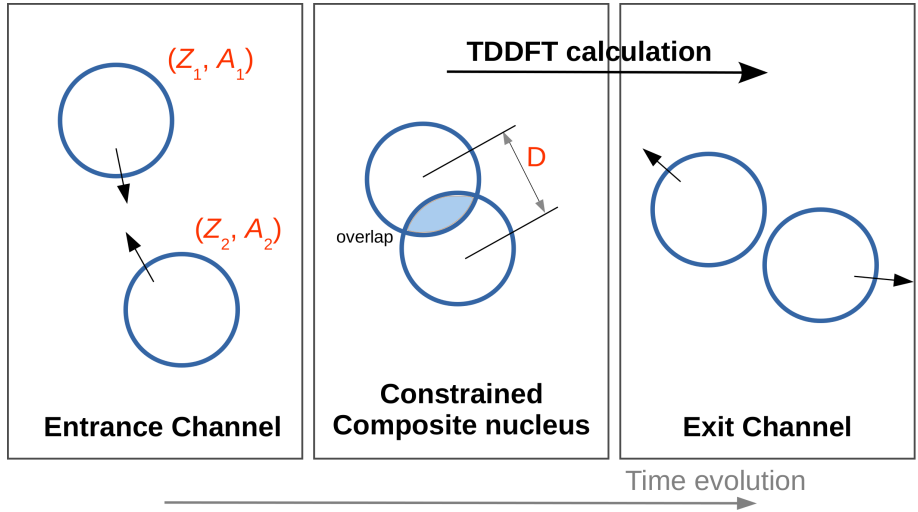}    
\caption{(Color online) Illustration of ``Composite-Nucleus Constrained TDDFT''.
Only the proton and mass numbers of colliding nuclei are given for the entrance channel (i.e., $Z_1$, $Z_2$, $A_1$ and $A_2$ in the left panel).
The two nuclei in the entrance channel might have relative velocity (shown as the arrows), but the velocities in the entrance channel are free parameters in this formalism.
There is a constraint on the composite nucleus in which the distance between the two center of mass is given (center panel).
The relative velocity of the two nuclei is set to exactly equal to zero for the composite nucleus, and they might have a certain overlap.
This constrained composite nucleus is the initial state of the TDDFT calculations.
The details of the exit channel such as the kind of emitted nuclei and the relative velocity between the nuclei are determined by the TDDFT calculations.
The conditions for the calculations are shown in red character.} 
\end{center}
\end{figure}

We present a formalism called ``Composite-Nucleus Constrained TDDFT'' to investigate the collision-fission dynamics.
In this article we mostly focus on the examination of charge equilibrium hypothesis of the collision-fission dynamics.
Collision-fission dynamics of a composite nucleus Thorium 240 is systematically investigated based on time-dependent density functional calculations employing SV-bas interaction parameter set~\cite{09klupfel}.

\section{Fission after collision}
\subsection{Charge equilibrium hypothesis}
A hypothesis, in which fission fragments produced from collision-fission
have almost the same $Z/N$-ratio to the composite nucleus, is assumed in
phenomenological treatments of studying the collision-fission events
(for the charge equilibration dynamics in collision-fission events, see \cite{88wahl, 07nishinaka, 11nishinaka}).
The reason is that quantum mechanical time-evolution including the nucleon
degrees of freedom is desired to determine the time evolution of charge distribution.
In this case $Z/N=90/150=0.60$ for the initial nucleus, and the point is to examine whether the $Z/N$-ratio of fission product is close to $0.60$ or not. 

\subsection{Composite-nucleus constrained TDDFT}
For a given energy and a given impact parameter, long-lived composite nuclei with certain excitation energies
(compared to the ground state) are possible to be produced. 
Let us take a composite nucleus of mass number $A$ and proton number $Z$. 
Consider binary collision-fission (Fig.~1):
\[  ^{A_1}Z_1 + ^{A_2}Z_2  ~\to~  ^AZ ~\to~   ^{A_1'}Z_1' + ^{A_2'}Z_2'   \]
where $A = A' = A_1 + A_2 = A_1' + A_2' = 240$ and $Z = Z_1 +Z_2 = Z_1' +Z_2' = 90$. 
First, choose the combination of the two nuclei at the entrance channel ($A_i$ and $Z_i$) to determine the heavy-ion reaction being
considered ($i=1,2$); six cases are taken as $(Z_1,A_1) = (10, 20)$,
$(20,40)$, $(30,60)$, $(30,80)$, $(40,80)$, and $(40,100)$.
Those states are obtained by the static density functional calculations.
Second, two states are put at a distance $D=$13.125~fm without giving any
velocities to the center of mass (center panel of Fig.~1), where the
diameter ($= 2R$) of
$A=240$ nucleus is roughly equal to 14.9~fm (cf. $R = 1.2 A^{1/3}$).
This corresponds to the constraint on the composite nucleus.
Third, the initial many-body wave function of the TDDFT calculations that is
eventually given as an orthogonalized single Slater determinant is given.
In fact a single Slater determinant consisting of two single wave functions of two different nuclei (two different Slater determinants) are orthogonalized before starting TDDFT calculations (cf. the Gram-Schmidt orthogonalization method), since the two single wave functions should have overlap. 
The initial many-body wave function physically simulates the excited composite nucleus in which the relative velocity of the original colliding nuclei is equal to zero.
Forth, the TDDFT calculation is carried out to see whether the fission appears or not and to see whether the fission fragments are in charge equilibrium or not.

\begin{figure}[t]   
\begin{center} \vspace{2.5mm}
\includegraphics[width=10.0cm]{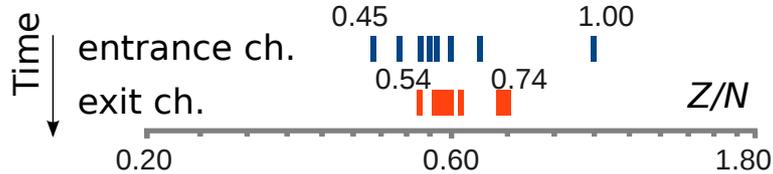} 
\caption{(Color online)
$Z/N$-ratio is plotted for entrance and exit channels (blue and red bars, respectively).
The logarithmic scale is adopted for the horizontal axis.
Duration time from the entrance to exit channels is roughly equal to 10$^{-20}$s.
The charge equilibrium corresponds to $Z/N = 90/150 =0.60$ in this case.
The maximal and minimal $Z/N$ values are also shown for reference.} 
\end{center}
\end{figure}

In this formalism two physical quantities are constrained: the shortest
distance of the two colliding nuclei (i.e., distance of the closest approach) and the kind of colliding nuclei at the entrance channel.
The most significant assumption included in this formalism is that no change is assumed to happen to the two colliding nuclei at the very early stage before the two colliding nuclei reach the shortest distance.
This formalism is originally proposed by Refs.~\cite{iwata-heinz}.
Certain difficulty in the standard TDDFT approach is expected to be overcome; indeed, the constrained composite nuclei sometimes hold the localization effect (e.g. clustering effect), which
can be lost in the mean-field type calculations such as the standard TDDFT.

\subsection{TDDFT Results}
The smaller one of the final products are $(Z,A-Z) = (7.6,10.5)$, $(18.0,24.6)$,
$(24.5,34.3)$, $(30,50)$, $(34.9,47.4)$ and $(38.2,61.3)$, which are produced from $(Z',A'-Z') = (10,10)$, $(20,20)$, $(30,30)$, $(30,50)$, $(40,40)$, and $(40,60)$, respectively.
One case results in fusion, and the others in binary fission.
Among five binary-fission cases, the masses of the lighter fragment decrease for three cases, which is pronounced to be a specific feature with the
fragmentation arising from lowly-excited composite nuclei \cite{10iwata-n}.

Figure 2 shows $Z/N$-ratios of the fragments for the entrance and exit channels.
This figure corresponds to a

 one-dimensional dynamical system projected in terms
of a physical quantity $Z/N$ in which the charge equilibrium ($Z/N = 0.60$)
plays a role of attractor.
Indeed the $Z/N$-range of the distribution becomes smaller from $0.55(=1.00-0.45)$ to $0.20(=0.74-0.54)$. 
Consequently it is confirmed in this specific setting that the charge equilibrium hypothesis is satisfied rather well.

\section{Concluding remark}
Charge equilibration hypothesis has been examined by means of a microscopic TDDFT framework, and this hypothesis has been confirmed to be satisfied well.
Alrhough, as for the TDDFT calculations, the charge equilibration dynamics was examined with nonzero initial relative-velocities, it is carried out with zero initial relative-velocities (i.e., zero TDDFT-initial velocities).
Despite researchers believed the possibility of the
non-negligible effects of relative velocity on the charge equilibration in old times, the present results suggest that such an effect is quite small. \\

\section*{Acknowlegements}
This work, which is based on the collaboration with Dr. Sophie Heinz (GSI Helmholtz Center for heavy-ion research) and Prof. Satoshi Chiba (Tokyo institute of technology), was supported by Research Laboratory for Nuclear Reactors, Tokyo Institute of Technology.
The author is grateful to the financial support by MEXT SPIRE and JICFuS. \\

\end{document}